\documentclass{mn2e}
\usepackage{natbib2,natbibmnfix}
\usepackage{astrojournals}
\usepackage{epsfig}
\usepackage{amssymb}
\usepackage{multirow}
\usepackage{multicol}
\usepackage{color}
\usepackage[varg]{txfonts}
\usepackage{url}

\def \mathbi#1{\textbf{\em #1}}
\def \msun{\rm M_\odot}

\begin{document}

\title[Binaries tear discs]{Tearing up the disc: misaligned accretion on to a binary}
\author[Nixon, King \& Price]{Chris~Nixon$^{1,2,4\star}$, Andrew~King$^2$ \& Daniel Price$^3$
\vspace{0.1in}\\
$^1$ JILA, University of Colorado \& NIST, Boulder CO 80309-0440, USA\\
$^2$ Department of Physics and Astronomy, University of Leicester, University Road, LE1 7RH Leicester, UK\\
$^3$ Monash Centre for Astrophysics (MoCA), School of Mathematical
Sciences, Monash University, Vic. 3800, Australia\\
$^4$ Einstein Fellow\\
$^\star$ chris.nixon@jila.colorado.edu}
\maketitle

\begin{abstract}
In a recent paper we have shown that the evolution of a misaligned disc around
a spinning black hole can result in tearing the disc into many distinct
planes. Tearing discs with random orientations produce direct dynamical
accretion on to the hole in $\approx 70\%$ of all cases. Here we examine the
evolution of a misaligned disc around a binary system. We show that these
discs are susceptible to tearing for almost all inclinations. We also show
that tearing of the disc can result in a significant acceleration of the disc
evolution and subsequent accretion on to the binary -- by factors up to $10^4$
times that of a coplanar prograde disc with otherwise identical
parameters. This provides a promising mechanism for driving mergers of
supermassive black hole (SMBH) binaries on timescales much shorter than a
Hubble time. Disc tearing also suggests new observational signatures of
accreting SMBH binaries, and other systems such as protostellar binaries.
\end{abstract}


\section{Introduction}
\label{intro}
Galaxy mergers are thought to be one of the main drivers of the coevolution of a galaxy and its supermassive black hole (SMBH). As most reasonably large galaxies contain an SMBH, mergers produce an SMBH binary. Dynamical friction with background stars allows the two SMBHs to sink to the centre of the merged galaxy and form a bound pair \citep{Begelmanetal1980}. This binary further interacts with stars by accreting or ejecting any which come close enough. This process is efficient at shrinking the binary to $\sim 1$ parsec, but at this point refilling the orbits which interact with the binary becomes too slow and the binary effectively stalls. This is the final parsec problem (\citealt{Begelmanetal1980}; \citealt{MM2001}).

There has been much focus on this problem in the literature. Gas should play an important role, as processes such as shocks can create low angular momentum gas which falls to the centre of the galaxy and forms a disc around the binary. However there is also progress with stellar dynamics; triaxial potentials repeatedly produce stellar orbits which interact with the binary, reducing the time taken to drive the merger \citep{Bercziketal2006}. The evolution of a triple SMBH system can also speed up the merger through the Kozai mechanism (\citealt{Blaesetal2002}; \citealt{Iwasawaetal2006}) or similarly an SMBH binary orbiting in a non-axisymmetric potential \citep{Iwasawaetal2011}.

Here we focus on the role of gas, specifically the evolution of a misaligned gas disc around a binary. Realistic, high--resolution simulations show that a coplanar prograde circumbinary disc is not efficient at shrinking the binary (\citealt{Cuadraetal2009}; \citealt{Lodatoetal2009}). This is simply because resonances between the binary and disc orbits are strong enough to hold the disc out, with little accretion. Therefore angular momentum is only extracted from the binary through the resonances with the disc orbits. This process is too slow to merge the binary in most cases, as any disc with enough mass to affect the binary sufficiently quickly is vulnerable to the gravitational instability, and so forms stars rather than shrinking the binary \citep{Lodatoetal2009}.

In contrast \cite{Nixonetal2011a} show that retrograde circumbinary discs are efficient in driving a binary merger as there are no orbital resonances in this case \citep{PP1977}. This allows the binary to accrete freely from the disc, and so gas with negative angular momentum is captured by the binary, reducing its orbital angular momentum and energy. \cite{Nixonetal2011a} show that gravitational interaction with a total retrograde--moving mass of $\sim M_2$ is required to drive the binary eccentricity to unity. Long before this is reached gravitational wave losses complete the coalescence. However there remains a problem. The mass of the circumbinary disc is limited by self--gravity to \citep{Pringle1981}
\begin{equation}
M_{\rm d} \lesssim \frac{H}{R} M_{\rm b}
\label{SG}
\end{equation}
where $M_{\rm d}$, $M_{\rm b}$ are the disc and total binary mass respectively, and $H/R$ is the disc angular semi--thickness. Any mass above this limit is likely to cause the disc to fragment and form stars, rather than accrete on to the binary \citep[see e.g.][]{Gammie2001}. Furthermore any mass added to the disc must be delivered at a rate many orders of magnitude below the Eddington limit to avoid fragmentation into stars \citep*{Kingetal2008}. Therefore unless $q=M_2/M_1 < H/R$, the merger requires more than one accretion event \citep*{Nixonetal2011b}. Typically black hole discs are radiatively efficient and therefore thin, with $H/R \sim 10^{-3}$ (\citealt{SS1973}; \citealt{CD1990}; \citealt{KP2007}). This may not hold close to the black hole innermost stable circular orbit where, at high accretion rates, radiation pressure dominates. However, for a circumbinary disc on scales approaching a parsec the disc is probably thin. Therefore typical mass ratios $0.01 \lesssim q \lesssim 1$ require many accretion episodes ($\sim q R/H$), some prograde and some retrograde, before the binary merges \citep*{Nixonetal2011b}. 

This picture is analogous to the chaotic accretion picture for black hole growth (\citealt{KP2006}; \citealt{KP2007}). Here a single black hole is subject to many small ($R\ll1$~pc, $M\lesssim \left(H/R\right)M_{\rm bh}$), randomly angled accretion episodes. In this case individual disc events are slightly more likely to lead to coalignment that counteralignment \citep{Kingetal2005} and so try to spinup the hole, but the retrograde events have a larger lever-arm and can efficiently spindown the hole. \cite{Kingetal2008} showed that the average black hole spin decreases with time (mass) but is around $a = 0.1-0.3$ for black holes of mass $10^6-10^{10}\msun$. There appears no reason to assume the gas dynamics differs on the scales of SMBH binaries $\lesssim 1$~pc, so we assume an equivalent model of small, randomly angled discs.

\cite{Nixonetal2011a} \& \cite*{Nixonetal2011b} restricted their attention to prograde and retrograde coplanar events as these seemed the only plausible outcomes of a misaligned disc. This reasoning was inferred from the analogous evolution of discs around spinning black holes, where (for $\alpha > H/R$; \citealt{PP1983}) the disc was expected to align quickly with the spin plane \citep{BP1975}. For a long time it was thought that only alignment (and not counter--alignment) occurred, but this resulted from the assumption in \cite{SF1996} of a disc with effectively infinite angular momentum. \cite{Kingetal2005} showed that when the magnitudes of the disc and hole angular momenta are finite, the disc counter--aligns whenever
\begin{equation}
\cos\theta < -\frac{J_{\rm d}}{2 J_{\rm h}}
\end{equation}
where $\theta$ is the angle between the disc and hole angular momentum $\mathbi{J}_{\rm d}$, $\mathbi{J}_{\rm h}$ with magnitudes $J_{\rm d}$ and $J_{\rm h}$. The analysis of \cite{Kingetal2005} also holds for circumbinary discs \citep*{Nixonetal2011b}, where the disc precession is now due to the non-spherical gravitational field of the binary (e.g. \citealt{Ivanovetal1999}; \citealt{Bateetal2000}; \citealt*{Nixonetal2011b}). The dominant effect of the potential comes from the zero-frequency (azimuthally symmetric $m = 0$)  term in the binary potential. This drives a differential precession in the disc. \cite*{Nixonetal2011b} suggested this would cause the disc to co-- or counter--align with the binary.

However, we now know of a third possible evolutionary track for these discs: breaking or tearing (\citealt{NK2012}; \citealt{Nixonetal2012b}). The implicit assumption before \cite{Nixonetal2012b} was that the disc viscosity would always be able to communicate the precession (Lense--Thirring in the black hole case; \citealt{LT1918}) rapidly enough to align the inner region of the disc to the spin, while a warp connected this to the still misaligned outer region. (For an isolated disc--hole system this warp propagates outwards until the whole disc is aligned.) But this assumption does not hold for typical disc--hole parameters. In \cite{Nixonetal2012b} we compared the magnitude of the torques induced by viscosity and precession, finding that near the black hole ($R < 350 R_{\rm g}$) the viscosity was not strong enough to hold the disc together, or more precisely that the precession is strong enough to tear the disc apart. This was confirmed in 3D numerical simulations which showed the disc breaking into distinct rings which precessed almost independently of the rest of the disc. In this regime, any disc inclined to the black hole spin by $45^\circ < \theta < 135^\circ$ ($\approx 70\%$ of randomly oriented events) is able to precess so that disc orbits become partially opposed, creating shocks which cancel angular momentum and so cause dynamical infall of gas \citep*[cf.][]{Nixonetal2012a}. We can see this by considering the angular momentum vector of the disc with Euler angles $\theta$ (tilt or zenith angle) and $\phi$ (twist or azimuth angle) \citep[e.g.][]{Pringle1996}
\begin{equation}
\mathbi{l}=\left(l_x,l_y,l_z\right) = \left(\cos\phi\sin\theta,\sin\phi\sin\theta,\cos\theta\right)
\end{equation}
where the binary angular momentum vector points in the $z$--direction. Now if we precess this vector by $\pi$ about the binary angular momentum vector, we simply change $\phi \rightarrow \phi+\pi$
\begin{eqnarray}
\mathbi{l}^\prime &=& \left(\cos\left(\phi+\pi\right)\sin\theta,\sin\left(\phi+\pi\right)\sin\theta,\cos\theta\right) \nonumber \\
        &=& \left(-\cos\phi\sin\theta,-\sin\phi\sin\theta,\cos\theta\right).
\end{eqnarray}
Now $\mathbi{l}\cdot \mathbi{l}^\prime = \cos\left(2\theta\right)$. Therefore the internal angle between these two rings of gas is $2\theta$. For any $\theta \ne 0$ angular momentum is reduced by the interaction of these two rings. To directly cancel angular momentum from gas orbits with opposed velocities we require the angle between them ($2\theta$) to be more than $90^\circ$, i.e. $45^\circ < \theta < 135^\circ$. The loss of angular momentum (and energy) allows the gas to fall to a smaller circularisation radius (Section 2 of \citealt*{Nixonetal2012a}).

Tearing the disc dramatically speeds up its evolution, by
circumventing the angular momentum barrier to accretion which would
otherwise require the gas to wait for viscosity to make it spiral
in. Gas rings draw on the binary
as a reservoir of angular momentum, in such a way that they are able
to cancel a significant fraction of their  own angular momentum.  
Here we explore whether such evolution is possible in a
circumbinary disc, and speculate how this evolution can help to speed
up the binary orbital evolution.

\section{Binary--disc tearing}
Following \cite{Nixonetal2012b} we compare the relevant torques. The viscous torque per unit area is (\citealt{LP1974}; \citealt{Franketal2002})
\begin{equation}
G_{\nu} = \frac{3\pi\nu\Sigma R^2 \Omega}{2\pi R H}
\end{equation}
and the precession torque per unit area induced by the binary is 
\begin{equation}
G_{\rm p} = \left|{\bf \Omega}_{\rm p} \times \mathbi{L}\right|,
\end{equation}
where the precession frequency is given (to first order) by (\citealt{Ivanovetal1999}; \citealt{Bateetal2000}; \citealt*{Nixonetal2011b})
\begin{equation}
{\bf \Omega}_{\rm p} = \frac{3}{4} \frac{M_2}{M_1+M_2} \frac{a^2}{R^2} \Omega \cos\theta.
\end{equation}
Therefore
\begin{equation}
\label{gp}
G_{\rm p} = \frac{3}{4} \frac{M_2}{M_1+M_2}a^2\Sigma \Omega^2 \cos\theta\sin\theta.
\end{equation}
To break the disc we require $G_{\nu} \lesssim G_{\rm p}$, i.e.
\begin{equation}
R_{\rm break} \lesssim \left(\frac{1}{4}\mu\left|\sin 2\theta\right|\frac{R}{H}\frac{1}{\alpha}\right)^{1/2} a,
\label{crit}
\end{equation}
where $\mu = M_2/(M_1+M_2)$ and we have used a \cite{SS1973} $\alpha$
with $\nu=\alpha c_{\rm s} H$ (for a discussion on viscosity see
Section~\ref{discussion}). Note that tearing is strongest for
$\theta=\pi/4,3\pi/4$ and there is no precession for $\theta =
0,\pi/2,\pi$ (as opposed to $\theta = 0,\pi$ in the Lense--Thirring case).

For typical parameters this gives
\begin{equation}
R_{\rm break} \lesssim 50\mu^{1/2}\left|\sin 2 \theta\right|^{1/2} \left(\frac{H/R}{10^{-3}}\right)^{-1/2} \left(\frac{\alpha}{0.1}\right)^{-1/2} a.
\label{rbreak}
\end{equation}
This suggests that tearing of circumbinary discs is inevitable near the binary, and can drag in significant amounts of gas on near--dynamical orbits. For example, a binary with separation $\sim 0.1$~pc with the parameters in Eq.~\ref{rbreak} could accrete gas from $\sim 5$~pc. We note that it is unlikely that a standard Shakura-Sunyaev disc extends to parsec scales, but instead suggest that gas on these scales can be significantly perturbed. This is likely to drive shocks, cancelling angular momentum and causing gas infall towards the binary. 

\section{Simulations}
\label{simulations}
We use the Smoothed Particle Hydrodynamics \citep[e.g.][]{Price2012} code
\textsc{phantom} (see e.g. \citealt{PF2010}; \citealt{LP2010}; \citealt*{Nixonetal2012a}; \citealt{Nixon2012})
to simulate the evolution of circumbinary discs. \textsc{phantom} has been
extensively benchmarked for performing simulations of warped discs
\citep{LP2010}, showing excellent agreement with the analytical treatment of
\cite{Ogilvie1999}. This is expected, as both treatments solve the
Navier--Stokes equations with an isotropic viscosity. The connections between
the viscosity coefficients derived by \cite{Ogilvie1999} therefore naturally
hold in our numerical treatment \citep{LP2010}.

\subsection{Setup}
The simulations use a disc viscosity with Shakura \& Sunyaev $\alpha \simeq
0.1$ \citep[][section 3.2.3]{LP2010} and a disc angular semi--thickness of
$H/R \simeq 0.01$ at the inner edge of the disc. Initially the disc has no
warp, and extends from an inner radius of $2a$ to an outer radius of $8a$,
with a surface density profile $\Sigma = \Sigma_0 (R/R_0)^{-p}$ and locally
isothermal sound speed profile $c_{\rm s} = c_{{\rm s},0} (R/R_0)^{-q}$, where
we have chosen $p=3/2$ and $q=3/4$ to achieve a uniformly resolved disc with a
constant $\alpha$ viscosity \citep{LP2007}. The disc is initially composed of
four million particles, which for this setup gives $\left<h\right>/H \approx
0.6$ \citep[cf.][]{LP2010}, where $\left<h\right>$ is the shell--averaged
smoothing length, such that $\alpha_{\rm AV} = 1.7$ corresponds to
$\alpha=0.1$. Given the strong supersonic (${\rm Mach} \lesssim 100$) shocks
present once the discs start to tear, we use $\beta_{\rm AV}=4$ to prevent
particle penetration \citep{PF2010}. For simplicity we consider an equal mass,
circular binary (we will study the effect of different mass ratios and
eccentricities in the future). The binary is represented by two Newtonian
point masses with accretion radii $0.05a$. The simulations differ only by the
relative inclination angle between the disc and the binary.

\subsection{Does the disc break?}
\begin{figure*}
  \begin{center}
    \includegraphics[angle=0,width=\textwidth]{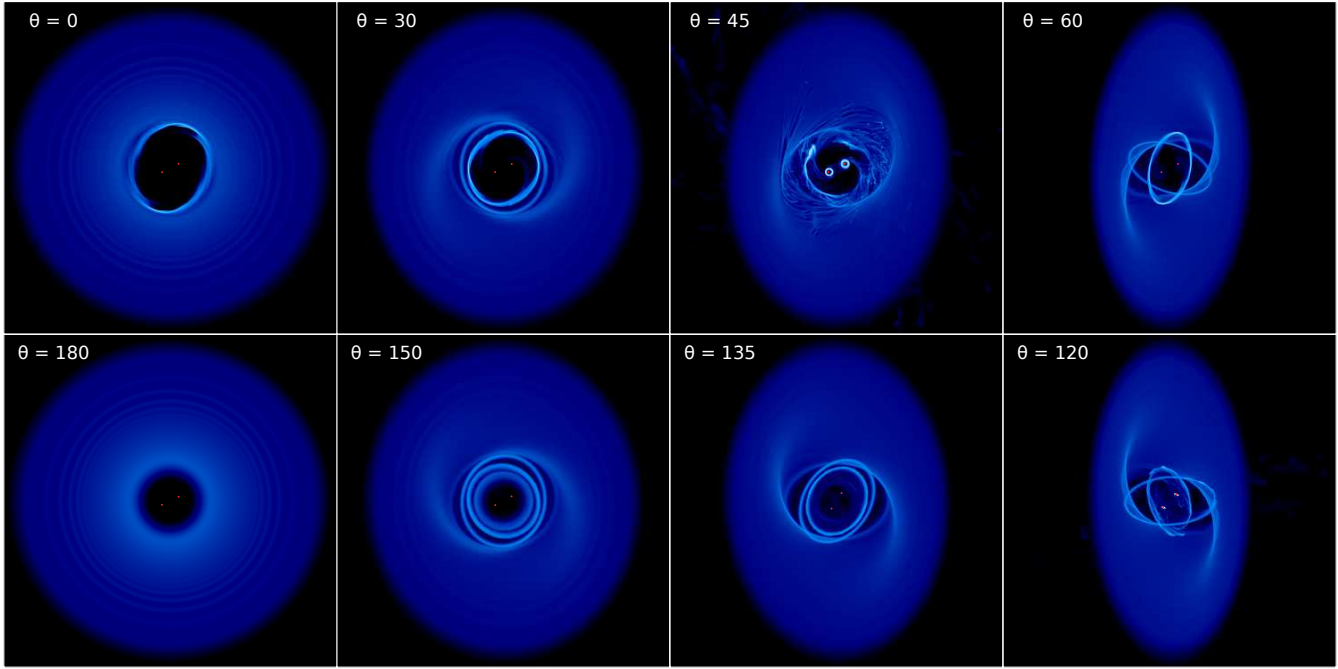}
    \caption{Column density plots of the simulations at $t=500$ ($\approx 80$ orbits of the binary) viewed along its axis. The top row shows discs prograde with respect to the binary; from left to right $\theta = 0,30,45,60$. The bottom row shows the corresponding retrograde discs with $\theta=180,150,135,120$. Fig.~\ref{comp2} shows the view along the binary plane. We note that although these snapshots are taken after the same number of binary orbits, this is not the same number of disc precession times, since this depends on the disc tilt (cf. Eq.~\ref{gp}). The maximum precession is achieved at $\theta=45^\circ$ \& $135^\circ$.}
    \label{comp}
  \end{center}
\end{figure*}

\begin{figure*}
  \begin{center}
    \includegraphics[angle=0,width=\textwidth]{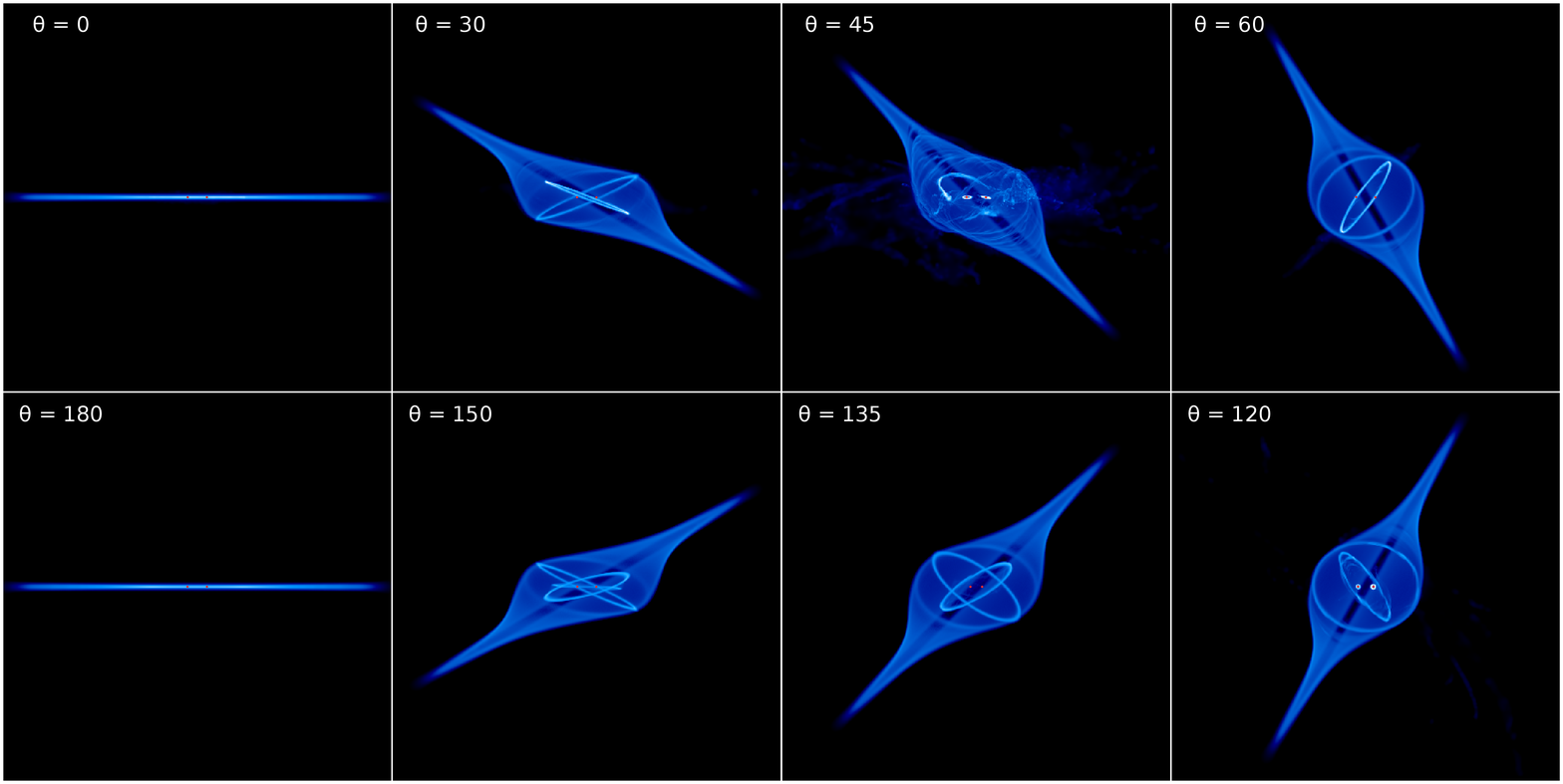}
    \caption{As for Fig.~\ref{comp}, but now viewed along the binary plane.} 
    \label{comp2}
  \end{center}
\end{figure*}

\begin{figure*}
  \begin{center}
    \includegraphics[angle=0,width=\textwidth]{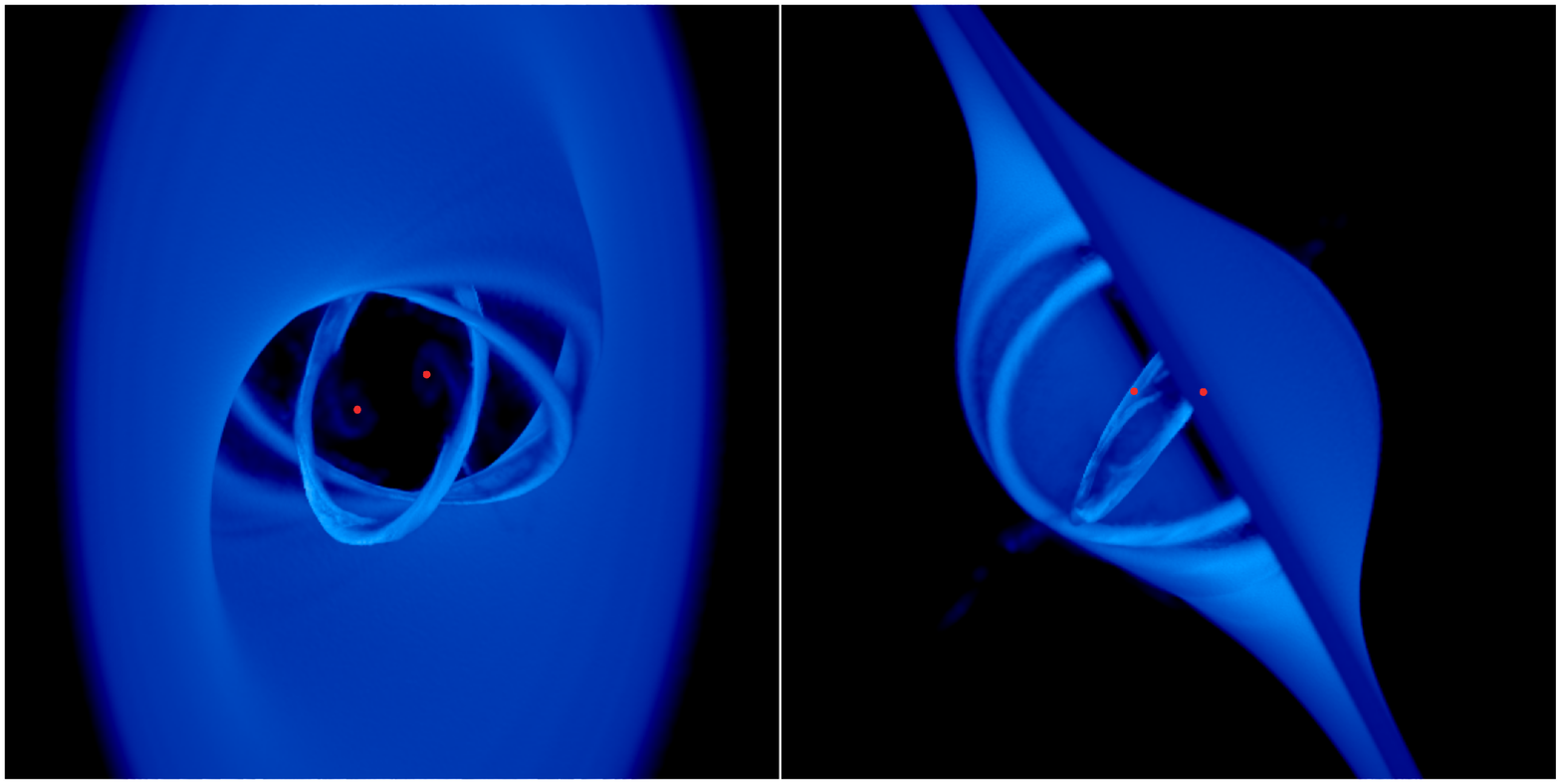}
    \caption{The 3D structure of the $\theta=45^\circ$ snapshot from Figs.~\ref{comp}~\&~\ref{comp2}.}
    \label{threeD}
  \end{center}
\end{figure*}

Yes. Figs.~\ref{comp}~\&~\ref{comp2} show snapshots of the simulations after 50 orbits of the binary for various inclinations. As expected the inclined discs precess around the binary. The expected breaking radius for the parameters in our simulation is given by
\begin{equation}
R_{\rm break} \lesssim 10 \left|\sin 2\theta\right|^{1/2} \left(\frac{\mu}{0.5}\right)^{1/2} \left(\frac{H/R}{10^{-2}}\right)^{-1/2} \left(\frac{\alpha}{0.1}\right)^{-1/2} a.
\end{equation}
As our initial disc extends from $2a \rightarrow 8a$ we expect the disc to break for any angle greater than a few degrees from $\theta=0$, $\pi/2$ \& $\pi$. This is supported by the simulations where only $\theta=0$ \& $\pi$ do not break (see Figs.~\ref{comp}~\&~\ref{comp2}). In Fig.~\ref{threeD} we show the 3D structure of the $\theta=45^\circ$ simulation.

\subsection{Prograde vs. Retrograde}
For a misaligned disc (with negligible angular momentum wrt the hole
spin) around a spinning black hole, there is symmetry about the spin
plane, so that discs with an inclination $\theta$ or $\pi-\theta$ produce similar evolution (see e.g. \citealt{Kingetal2005} for a discussion). However this is not true in the binary case. Resonances between the binary and disc orbits occur in prograde discs only 
($\theta < 90^\circ$; \citealt{PP1977}; \citealt{Nixonetal2011a}), and allow the binary to transfer angular momentum to the disc orbits, rather than just a precession. 

This asymmetry produces a distinct difference between prograde and retrograde circumbinary discs \citep[cf.][]{Nixonetal2011a}. The lack of resonances in the retrograde simulations mean that the evolution is comparatively simple. Each gas orbit precesses around the binary, remaining near--circular unless it interacts strongly with another gas orbit. This is similar to the tearing evolution of a misaligned disc around a spinning black hole \citep{Nixonetal2012b}. 

\begin{figure}
  \begin{center}
    \includegraphics[angle=0,width=\columnwidth]{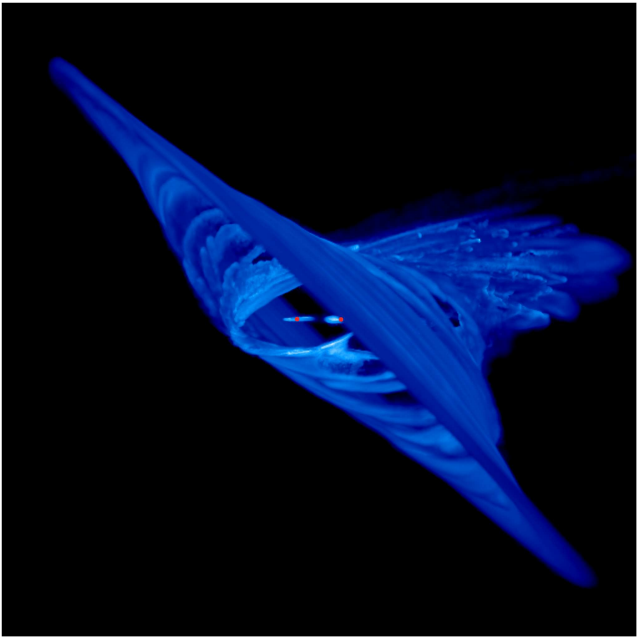}
    \caption{3D structure of the $45^\circ$ simulation at $t = 700 \approx 110$ orbits of the binary. In a prograde tearing disc resonances drive the innermost circumbinary gas significantly eccentric.}
    \label{resonate}
  \end{center}
\end{figure}

In contrast, in prograde discs there is additional transfer of angular momentum through resonances. In a prograde coplanar ($\theta=0^\circ$) disc, the extra angular momentum given to the disc gas is dissipated rapidly, so that (except very close to the binary) the disc remains near--circular, but moves out slightly (at some point finding a balance with viscous torques). However if most of the disc is tilted, but the inner parts are aligned to the binary plane, the angular momentum added to the inner parts is no longer always efficiently dissipated in the disc -- as there may be no gas readily available to interact with the gas resonating with the binary orbit. Instead the disc gas can become more eccentric, see Fig.~\ref{resonate}. This may change the observational appearance of circumbinary discs.

If the disc is inclined by $45^\circ < \theta < 135^\circ$ disc
tearing produces orbits that cancel angular momentum and so leads to
dynamical gas infall. If the disc is retrograde this gas is accreted
directly by the binary. If the disc is  prograde, this gas is either
accreted, or dynamically interacts with the binary and is kicked out
to large radii. This gas ejection is clearly sensitive to the accretion radii of
the primary and secondary, and numerical simulation of it requires
high resolution in the interaction region, so we will return to this
in the future with a more thorough investigation. This process could
be an important ingredient in solving the last parsec problem, as this
gas could carry away a significant fraction of the binary orbital
angular momentum. The ejected gas may shock and interact with gas in
the host galaxy, possibly producing an observational signature, but
may also fall back to interact again with the binary.

\subsection{Accretion rate}
It is clear that tearing the disc significantly increases the accretion rate on to the binary compared to the corresponding coplanar prograde disc. Fig.~\ref{accretion} gives the relative accretion rates. Accretion in the simulations that tear is strongly enhanced -- by factors up to $10^4$ compared with the coplanar prograde simulation. Prograde discs accrete less than the corresponding retrograde discs, simply because resonances hold the gas out in the prograde cases.

We report the percentage of the disc accreted by the binary in Table~\ref{percentages}. This data is taken at $t=500$, which corresponds to $\sim 80$ binary orbits.
\begin{table}
  \begin{tabular}{|l|c|c|c|c|c|c|c|c|}
    \hline
    inclination & 0    & 30   & 45  & 60  & 120 & 135 & 150 & 180  \\ \hline
    \% accreted & 0.02 & 0.08 & 1.2 & 0.1 & 6.0 & 5.0 & 0.6 & 0.1 \\
    \hline
  \end{tabular}
  \caption{Percentage of the disc accreted for different inclination angles. This data is taken at $t=500$, after approximately $80$ binary orbits.}
  \label{percentages}
\end{table}
Note that some of the accretion rates are not resolved in these simulations (see Fig~\ref{accretion}). However, this does highlight that accretion through coplanar discs ($\theta=0,180$) is significantly faster if the disc is retrograde. Thus the timescale $M_2/{\dot M}$ suggested by \cite{Nixonetal2012a} for the binary merger is significantly shorter than the accretion time for prograde discs. If the disc is torn up this timescale is again reduced by orders of magnitude.

\section{Discussion}
\label{discussion}
\subsection{Assumptions}
We have shown that almost all SMBH circumbinary discs are likely to tear (see Eq.~\ref{rbreak}), and confirmed this using 3D simulations. These results rest on three main assumptions, which we discuss here.

The first assumption is that accretion events on to an SMBH binary are likely to be randomly oriented, and not all perfectly aligned to the binary plane. This is similar to the chaotic accretion scenario for growing SMBH (\citealt{KP2006}; \citealt{KP2007}; \citealt{Kingetal2008}). Accretion is likely to be chaotic because the scale of the binary ($\lesssim 1$~pc) is many orders of magnitude smaller than the scale of the galaxy region feeding it. Processes internal to the galaxy, such as star formation, supernovae and stellar winds can randomize the angular momentum of infalling gas, so there is no reason to expect accretion events to be initially aligned with the binary.

The second assumption concerns the disc viscosity. We have assumed
that this can be adequately represented by an isotropic \cite{SS1973}
$\alpha$ viscosity. This is what is modelled in our SPH code (see
\citealt{LP2010} for details) and what is used in Eq.~\ref{crit}. In
reality the viscosity is probably driven by turbulence of some
form. For typical black hole discs this is expected to be the
magnetorotational instability (MRI) \citep{BH1991}, while at $\sim$
parsec scales turbulence induced by self--gravity is also likely to be
important for discs with enough mass. We have assumed a high viscosity
($\alpha = 0.1$) and still find that the disc readily breaks, so it
appears that a realistic $\alpha$ cannot prevent tearing if the
viscosity is isotropic. In this sense our simulations are conservative
with respect to tearing. The only way the disc viscosity might resist
tearing is if it is significantly anisotropic, in the sense of
favouring the vertical viscosity. This cannot be ruled out, but seems
unlikely. As pointed out by \cite{Pringle1992} azimuthal shear is
secular and vertical shear is oscillatory. This suggests that any
viscosity anisotropy might instead {\it favour} tearing.
 
Finally we have assumed an isothermal equation of state for the disc gas. This is equivalent to assuming the gas radiates away any heating instantly. In reality shocks heat the gas and this must be radiated away on a cooling time. As long as the cooling time is shorter than the time taken for the disc to precess our assumption holds. If however the gas is unable to cool, the excess heat increases the thickness of the disc and might in principle enable the disc to resist tearing (cf. Equation~\ref{crit}). But to affect the tearing radius by an order of magnitude, the disc must thicken by 2 orders of magnitude -- and thus decrease the viscous time by 4 orders of magnitude, i.e. increase the accretion rate by this amount. For counterrotating rings, the accretion rate is somewhat insensitive to the equation of state; if the heat is retained in the gas the flow becomes more chaotic, causing more mixing and a similar level of accretion \citep{Nixonetal2012a}. We also note that our simulations have $H/R = 0.01$, rather than $10^{-3}$ we expect for such discs. This is simply because $H/R \approx 0.01$ is the limit of what we can currently resolve. Therefore again, if the isothermal assumption holds, our simulations are conservative with regards to tearing. If the disc gas cannot cool, it must drive enhanced accretion to have any chance of resisting tearing. We will explore this alternative possibility in the future.
\begin{figure}
  \begin{center}
    \includegraphics[angle=0,width=\columnwidth]{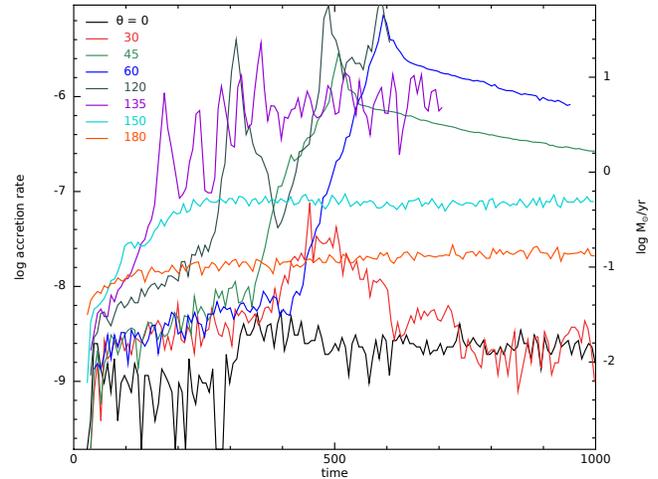}
    \caption{Accretion rates for the simulations. The accretion rate is
      calculated in time bins of width one binary orbital time ($2\pi$), with
      the time axis in units of the binary dynamical time. The accretion rate
      is in arbitrary units (binary mass per dynamical time). An accretion
      rate $\sim 10^{-10}$ corresponds to approximately 1 particle per binary
      orbit. Therefore anything less than a few $\times 10^{-9}$ is not
      resolved. The simulations with $\theta=0^\circ$ \& $30^\circ$ fall into
      this category as the resonances hold most of the gas out, preventing any
      noticeable accretion. However if the disc is retrograde or tears then
      the accretion rates are much higher. Our simulations suggest that for
      these parameters, circumbinary discs which tear can have accretion rates
      up to $10^{4}$ times those of coplanar prograde discs. If we scale the
      simulations, see RHS axis label, with a total binary mass $2\times
      10^8\msun$, disc mass $10^6\msun$ and binary separation $0.1$~pc, then
      $10^{-9}$ is approximately $0.006\msun/{\rm yr}$, whereas the highest
      accretion rates ($\sim 10^{-5}$) are up to $60\msun/{\rm yr}$.}
    \label{accretion}
  \end{center}
\end{figure}

\subsection{Wave-like propagation of warps}
To derive Eq.~\ref{crit} we have assumed that viscosity provides the dominant internal disc torque. However, discs can communicate angular momentum by propagating waves (e.g. \citealt{LP1993}; \citealt{PL1995}; \citealt{LO2000}). For wave-like disturbances to dominate, the disc must be both nearly Keplerian and nearly inviscid, specifically \cite[e.g.][]{Ogilvie1999}
\begin{equation}
\left|\frac{\Omega^2-\kappa^2}{\Omega^2}\right| \lesssim H/R ~~~~~{\rm and}~~~~~\alpha\lesssim H/R
\end{equation}
where $\Omega$ is the angular velocity and $\kappa$ is the epicyclic frequency.

For our simulations neither of these criteria are satisfied. However, if the mass ratio of the binary is extreme the disc orbits are closer to Keplerian, and if the disc is substantially thicker, wave propagation becomes important. \cite{Larwoodetal1996} \& \cite{LP1997} have performed simulations of circumbinary discs covering a wide range of parameters. Most of their simulations have unequal mass ratios and thick discs, which lead to our estimate of the breaking radius (\ref{crit}) being inside the inner edge of their discs. However, equation (\ref{crit}) suggests models 12 \& 13 of \cite{LP1997} should break. This behaviour is seen in Model 13, but not as strongly as Eq.~\ref{crit} would suggest, while Model 12 shows no sign of breaking. Therefore it is likely that in these cases the viscous torque used to calculate Eq.~\ref{crit} is inadequate. This is not surprising as the calculation leading to (\ref{crit}) neglects the pressure effects which appear to dominate in the simulations of \cite{LP1997}. Similarly the calculation is insensitive to the vertical viscosity in accretion discs \citep[see e.g.][for a discussion]{Pringle1992}. Therefore we caution that this estimate is purely a guide to determine if disc breaking/tearing is likely to occur. However it does appear that breaking of thick wave-like discs is possible, and demonstrated by the numerical simulations of \cite{LP1997}. We suggest a plausible criterion for breaking wave-like discs in Appendix~\ref{appA}. 

We also note that similar simulations have been performed by \cite{FN2010}. Here the disc is around the primary star, rather than the binary, but the physical process is the same. We shall report in more detail in a future paper, but the simulations of \cite{FN2010} appear in broad agreement with the equivalent estimate for breaking of a circumprimary disc. The agreement with the radius defined from a viscous model is better in this case as these simulations often used $\alpha > H/R$, and so the pressure effects are reduced. We note that the simulations of \cite{FN2010}, which also report disc breaking, use a grid--based numerical method ({\sc nirvana}; \citealt{ZY1997}).

\section{Conclusions}

In \cite{Nixonetal2012b} we found that a misaligned disc around a spinning black hole could be torn into distinct rings which precess almost independently. For most randomly--oriented accretion events this leads to direct cancellation of angular momentum and so to dynamical accretion of gas. Here we have shown that misaligned circumbinary discs also tear (see Figs.~\ref{comp} \& \ref{comp2}). Fig.~\ref{accretion} shows that tearing can produce accretion rates which are $\sim 10^4$ times higher than if the same disc were prograde and coplanar. In prograde discs, tearing and subsequent dynamical infall of gas counteracts to some degree the negative effect of resonances on accretion. Retrograde circumbinary discs do not suffer from resonances and so their accretion rates are higher than the corresponding prograde circumbinary disc.

We have also found two interesting consequences of tearing in circumbinary
discs which may be observable. The large eccentricities in tilted prograde
discs may give an observable signature (shocks or
star formation) when the gas reaches high densities at pericentre. Ejection of
gas by the binary may also lead to observable outflows from a circumbinary
disc system. We will return to these effects in a future
investigation.

Eq.~\ref{rbreak} suggests that for realistic parameters, tearing can occur in discs at radii up to 50 times the binary separation. For SMBH binaries with separation $\sim 0.1$~pc this is approaching the gravitational influence radius of the binary, making this a promising mechanism for dragging gas on to the binary. On scales larger than this, the gravitational potential of the wider galaxy must be taken into account. The radius for tearing (\ref{rbreak}) is reduced if the binary has an unequal mass ratio. However, the dependence is modest (Eq~\ref{crit}: $R_{\rm break}/a \propto \mu^{1/2}$), so to prevent tearing entirely the mass ratio must be extreme (e.g. $\mu \lesssim 0.001$). This suggests that binaries composed of a star and a planet, or the extreme end of SMBH binaries, are unlikely to tear external discs. However stellar binaries and typical SMBH binaries are quite capable of tearing up a sufficiently inclined circumbinary disc.

\begin{figure*}
  \begin{center}
    \includegraphics[angle=0,width=\textwidth]{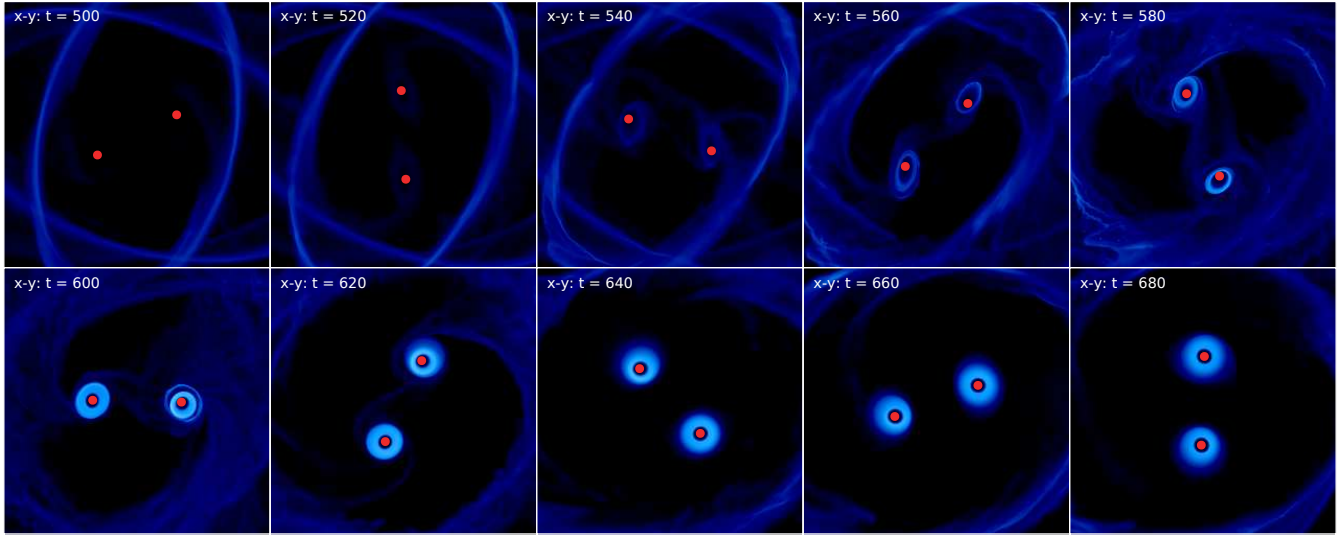}
    \caption{Zoomed in view of the $\theta=60^{\circ}$ simulation while two rings are interacting. Only particles close to the binary ($r < 2.5$) are plotted, showing the sub-discs form around the primary and secondary. Initially the discs are significantly misaligned to the binary; over time they align to the binary plane. The accretion during the formation of these discs can be seen in Fig~\ref{accretion}.}
    \label{subdiscs}
  \end{center}
\end{figure*}

\begin{figure*}
  \begin{center}
    \includegraphics[angle=0,width=\textwidth]{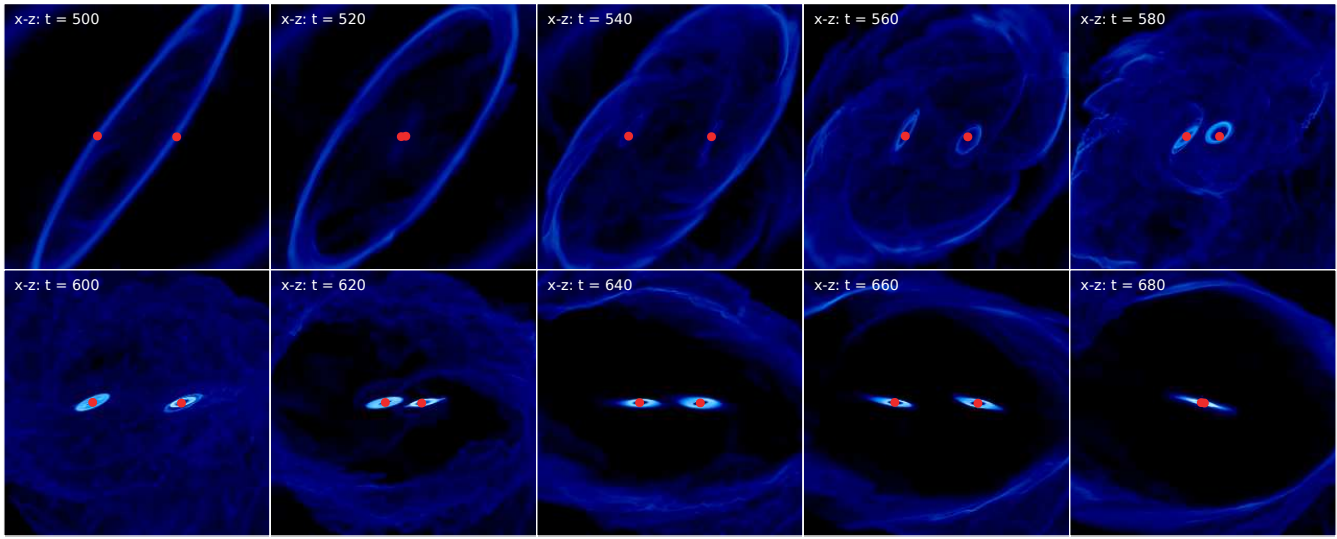}
    \caption{As for Fig.~\ref{subdiscs}, but now viewed along the binary plane.}
    \label{subdiscsb}
  \end{center}
\end{figure*}

Tearing circumbinary discs produce dynamical infall of gas. This gas interacts with the binary and often forms circumprimary and circumsecondary discs. In Figs.~\ref{subdiscs} \& \ref{subdiscsb} we show the formation of these discs in the $\theta=60^{\circ}$ simulation. Initially the discs form with a significant misalignment to the binary orbit. The vertical structure of these smaller discs is below our resolution limit, effectively increasing their viscosity and removing the possibility of tearing in them, but we will return to this question in a future publication. If tearing does occur, it may allow a cascade of instabilities to funnel gas from $>$~pc scales on to the black hole by repeated precession, angular momentum cancellation and direct infall.

For retrograde discs \cite{Nixonetal2011a} show that the timescale for merging the binary is $\sim M_2/{\dot M}$. Here we have shown that tearing can increase ${\dot M}$  by many orders of magnitude in a circumbinary disc. This could have a direct impact on the merger timescale. However when we consider the constraints of disc self--gravity (Eq.~\ref{SG}) it is clear that a single accretion event does not have enough mass to drive the merger. Therefore it is not straightforward to evaluate the merger timescale. We will investigate the effect of multiple randomly oriented accretion events in the future.

If the gas supply from the galaxy is slower than accretion through a prograde coplanar disc, then tearing cannot accelerate the merger as the timescale is limited by the mass supply rate. However, if the galaxy can supply mass above this rate, the bottleneck becomes how fast the gas can be transported through the disc. Tearing can speed this up by a factor up to $10^4$. Using the example in Fig.~\ref{accretion} ($10^6\msun$ disc  at $0.1$~pc from a $2\times 10^8\msun$ binary), the accretion rate is boosted from $\sim 0.006$ to $\sim 60\msun/{\rm yr}$. In a gas--rich merger, it seems likely that gas is supplied to the central regions fast enough for the disc evolution to dominate the merger timescale.

We also note that the maximum sustainable accretion rate in a
critically self--gravitating disc is a strong function of radius (see
\citealt{Levin2007}; \citealt{Harsonoetal2011}). This rate decreases
by up to 3 orders of magnitude where the disc goes from optically
thick to optically thin (see Fig.~1 of Harsono et al. 2011). Tearing
the disc can boost the accretion rate enough to avoid this
gravitational catastrophe, but even with near--dynamical gas infall
the disc mass is still limited to a small fraction ($\lesssim H/R$) of
the binary mass. However if the disc cannot radiate away fast enough
the heat provided by tearing, it must thicken substantially (see
Eq.~\ref{rbreak}). This may allow an otherwise gravitationally
unstable disc to survive and drive the binary merger in a single
event.

Finally, tearing circumbinary discs may also occur in other systems, such as protostellar binaries. If star formation is sufficiently chaotic \citep[e.g.][]{Bateetal2010} then misaligned circumbinary discs are likely to form. Tearing these discs may promote a period of rapid, variable accretion on to the protostars at early times. This could also explain any misalignments between the stellar and binary rotation axes, or misalignments between the binary orbit and any planets in the system.

\section*{Acknowledgments} 
We thank the referee for useful suggestions. Support for this work was
provided by NASA through the Einstein Fellowship Program, grant
PF2-130098. Research in theoretical astrophysics at Leicester is supported by
an STFC Consolidated Grant. We used {\sc splash} \citep{Price2007} for the
visualization. The calculations for this paper were performed on the
Complexity node of the DiRAC2 HPC Facility which is jointly funded by STFC,
the department of Business Innovation and Skills and the University of
Leicester.

\bibliographystyle{mn2e} 
\bibliography{nixon}

\appendix
\section{Tearing wave-like discs}
\label{appA}
Here we provide an estimate of where a pressure--dominated, wave--like disc may break. In discs with $\alpha \ll H/R$ warping disturbances propagates as waves \citep{PP1983} travelling at roughly half the local sound speed (\citealt{PL1995}; \citealt{Pringle1999}; \citealt{Lubowetal2002}). Therefore a simple estimate of the wave communication time in a disc is given by
\begin{equation}
t_{\rm w} \sim \frac{2R}{c_{\rm s}}.
\end{equation}

We compare this to the precession timescale to derive a criterion for breaking wave--like discs
\begin{equation}
\frac{1}{\Omega_{\rm p}} \lesssim \frac{2R}{c_{\rm s}} = 2 \frac{R}{H}\frac{1}{\Omega}.
\end{equation}
For the circumbinary discs considered above this gives
\begin{equation}
R \lesssim \left(\frac{3}{4}\mu\left|\sin2\theta\right|\frac{R}{H}\right)^{1/2}a
\end{equation}
whereas for the black hole accretion discs discussed in \cite{Nixonetal2012b} this is
\begin{equation}
R \lesssim \left(4a\left|\sin\theta\right|\frac{R}{H}\right)^{2/3} R_{\rm g}.
\end{equation}

We caution that these estimates make the assumption that the disc is inviscid. Damping of these waves occurs on a timescale $\sim \left(1/\alpha\right)t_{\rm dyn}$ \citep[e.g.][]{Lubowetal2002}. Therefore if $\alpha$ is $\sim 0.1-0.4$ as expected for fully ionised black hole discs \citep[e.g.][]{Kingetal2007} the waves may damp after only a small number of dynamical times. Therefore the maximum distance the wave can communicate is limited to $R_{\rm w}$ where
\begin{equation}
\frac{2R_{\rm w}}{c_{\rm s}} \approx \frac{1}{\alpha\Omega}
\end{equation}
and therefore
\begin{equation}
R_{\rm w} \approx \frac{H}{2\alpha}.
\end{equation}

Also, these calculations rely on a linear understanding of warp wave propagation. The full nonlinear effects are yet to be explored in this case (cf. \citealt{Gammieetal2000}; \citealt{Ogilvie2006}). We note that these calculations probably apply best to protostellar discs, where warps typically propagate as waves (e.g. \citealt{TB1993}; \citealt{PT1995}; \citealt{LO2000}; \citealt{NP2010}).

\end{document}